# Some new insights into information decomposition in complex systems based on common information

Pradeep Kr. Banerjee [1]

[1]Indian Institute of Technology Kharagpur, e-mail: pradeep.banerjee@gmail.com

September 18, 2014

**Abstract:** We take a closer look at the structure of bivariate dependency induced by a pair of predictor random variables ($X_1$, $X_2$) trying to synergistically, redundantly or uniquely encode a target random variable $Y$. We evaluate a recently proposed measure of redundancy based on the Gács and Körner's common information (Griffith *et al.*, *Entropy* 2014, 16, 1985–2000) and show that the measure, in spite of its elegance is degenerate for most non-trivial distributions. We show that Wyner's common information also fails to capture the notion of redundancy as it violates an intuitive monotonically non-increasing property. We identify a set of conditions when a conditional version of Gács and Körner's common information is an ideal measure of unique information. Finally, we show how the notions of approximately sufficient statistics and conditional information bottleneck can be used to quantify unique information.

## 1. Introduction

We take a closer look at the structure of bivariate dependency induced by a pair of predictor random variables (RVs) ($X_1$, $X_2$) trying to encode a target RV $Y$. The information that the pair ($X_1$, $X_2$) contains about the target $Y$ can have aspects of redundant information (contained identically in both $X_1$ and $X_2$), of unique information (contained exclusively in either $X_1$ or $X_2$), and of synergistic information (contained only in the joint RV ($X_1$, $X_2$)). For the general case of $K$ predictors, Williams and Beer [1] proposed one such decomposition called the partial information (PI) decomposition to specify how the total information about the target is shared across the singleton predictors and their overlapping or disjoint coalitions. However, effecting a non-negative decomposition is known to be a surprisingly difficult problem even for the case of $K = 3$ [2]. In particular, it is not always possible to attribute operational significance to all the atoms induced by the decomposition. What operational questions should an ideal measure of redundant or unique information answer? In this paper, we explore the bivariate case and demonstrate the richness of this question through the lens of network information theory.

We briefly motivate the problem with reference to several applications where information-theoretic notions of synergy and redundancy are deemed useful. One of the main challenges in computational neuroscience is to quantify the neural code, i.e., how well neural populations encode sensory information and what is the fidelity of such a representation [3], [4], [5], [6]. Single neurons are not very informative in that they do not encode the stimulus well and what counts is the activity of ensembles of neurons. In the context of neural coding, compound events in neural spike patterns may jointly carry far more information than that is carried independently by its parts, e.g., two spikes close together in time may carry far more information than the aggregate of the individual spikes, thus demonstrating synergy in the code [7]. Similarly, in computational genetics, cellular pathways are highly cooperative and diseases such as cancer can be better analyzed in terms of the synergy among multiple interacting genes, i.e., in terms of the purely synergistic, as opposed to independent nature of their contributions towards a phenotype [8]. The common objective of all these studies is to identify how the total informational load induced by the stimulus is shared amongst participating coalitions of a set of component sources (neurons, genes, etc.). Further motivating examples for studying information decomposition abound in distributed control [9] and adversarial settings like game theory [2], where notions of common knowledge shared between agents are used to describe epistemic states.

The organization of the paper is as follows. In Section 2, we discuss the existing notions of common information (CI). In Section 3, we introduce the PI decomposition framework [1] and evaluate a recently proposed measure of redundancy [10] based on the Gács and Körner's CI [13]. In Section 4, we briefly explore the subtleties in defining a combinatorial dual of the Gács and Körner's CI. We identify a set of conditions when a conditional version of Gács and Körner's CI is an ideal measure of unique information. Finally, we show how a modified framework of the information bottleneck principle [29] can be used to quantify unique information.

## 2. Common information measures

We use lowercase letters $x, y$, etc. to denote values of RVs and uppercase letters $X, Y$, etc. to denote RVs. $X^n$ denotes the sequence $(X_1, \ldots, X_n)$. The distribution (pmf) of the discrete RV $X$ is denoted by $p_X(x) = \Pr\{X = x\}, x \in \mathcal{X}$. We write $p_X(x)$ as $p(x)$ when there is no confusion. The entropy of $X$ is defined as $H(X) \triangleq \sum_{x \in \mathcal{X}} -p(x) \log p(x)$. For $(X, Y) \sim p(x, y)$, the Shannon mutual information between $X$ and $Y$ is defined as $I(X, Y) \triangleq H(X) + H(Y) - H(XY) = \sum_{(x,y) \in \mathcal{X} \times \mathcal{Y}} p(x, y) \log \frac{p(x,y)}{p(x).p(y)}$. $I(X; Y)$ is the most frequently used notion of common information (CI) and quantifies the descriptive savings in communication rate if the receiver has some prior side information about the messages being conveyed. Depending on the operational questions it answers, there are at least two other notions of CI, due to Gács and Körner [13] and Wyner [15]. Each of these notions appears as solutions to asymptotic formulations of some important information processing task.

*2.1. Information structures and the lattice of information σ-algebras*

The earliest ideas of representing information by a partition of the sample space dates back to Shannon [17]. A partition of a set $\mathcal{X}$ is a division of $\mathcal{X}$ into non-empty, disjoint subsets s.t. their union gives back the set $\mathcal{X}$. Given an underlying probability space, there exists a natural one-to-one mapping between sample-space partitions and σ-algebras. This implies we can partition the set of all RVs into disjoint equivalence classes, called *information elements*, s.t., all RVs within a given class are *informationally equivalent* [17], [18]. Shannon then defined a relation of inclusion between two such information elements: we say that $Y$ ($X$) is *informationally richer* (*poorer*) than $X$ ($Y$) if $Y \succcurlyeq X$ $\triangleq H(X|Y) = 0$. We write $X \cong Y$ to denote that $X$ and $Y$ are *informationally equivalent*. This naturally induces a partial order on the information elements and to keep track of the inclusion relations, Shannon defined the *information lattice* as a set of information elements closed under the sum (*join*) and product (*meet*) operations. The join of two RVs, $X(\in \mathcal{X})$ and $Y(\in \mathcal{Y})$, denoted by $Z = X \vee Y$ is simply the joint RV $Z = (X, Y)$, and is called the *total information* of both $X$ and $Y$. Similar, the meet of two RVs $X$ and $Y$, denoted by $Q = X \wedge Y$ is the largest RV $Q$ s.t. $Q \preccurlyeq X$, $Q \preccurlyeq Y$, i.e., $Q$ is the poorest among all information elements that are richer than both $X$ and $Y$. Shannon called $Q = X \wedge Y$ as the *common RV* of $X$ and $Y$. The join and meet operations are associative and commutative, but in general, information lattices are neither distributive, nor even modular [17]. Figure 1(a) shows an example of an information lattice [17] where $X$, $Y$ and $Z$ are three independent RVs. Figure 1(b) shows an example of a XOR lattice, $Z = \text{XOR}(X, Y)$, where the triple $(X, Y, Z)$ are pairwise independent. However, taking any pair uniquely determines the third one, i.e. sum of any two represents the total information in the system. The XOR lattice exemplifies the non-distributive nature of the information lattice [17], since $(Z \wedge Y) \vee (Z \wedge X) = 0$, whereas $Z \wedge (X \vee Y) = Z \neq 0$.

**Figure 1. (a)** Information lattice for three independent RVs $(X, Y, Z)$. **(b)** The XOR information lattice.

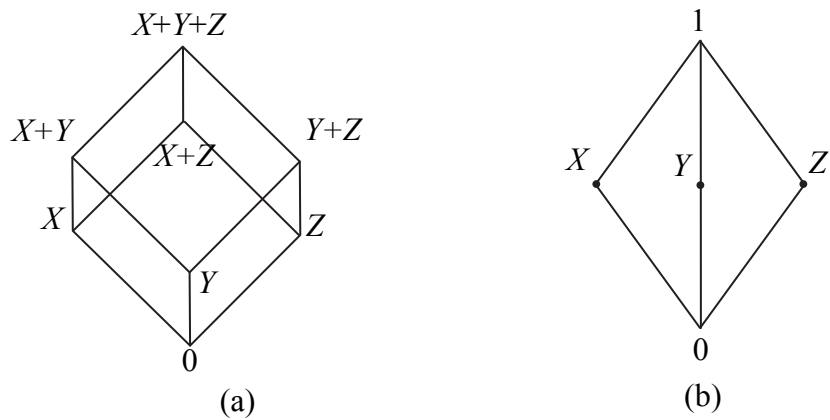

*2.2. Gács and Körner Common Information*

Almost two decades following Shannon's work [17], Gács and Körner's [13] independently proposed and studied in detail the notion of common information. For characterizing common information, Gács and Körner specified the maximum common function of $X$ and $Y$, mcf$(X,Y)$ as separate functions of $X$ and $Y$ that agree with probability 1, s.t. any other common function of $X$ and $Y$ is a function of mcf$(X,Y)$. Let $\{X_i, Y_i\}_{i=1}^{\infty}$ be i.i.d. copies of the pair $(X,Y) \sim p(x,y)$. Let $(f^n, g^n)$ be a pair of deterministic mappings and let $Q = f^n(X^n)$ and $Q' = g^n(Y^n)$. Let $\varepsilon_n = \Pr\{Q \neq Q'\}$, $\rho_n = \frac{1}{n}H(Q)$, and $(f^n, g^n)_{n=1}^{\infty}$ be a sequence of such mappings for which $\lim_{n\to\infty} \varepsilon_n = 0$. For this sequence, let $\rho_\infty = \limsup_{n\to\infty} \rho_n$. Thus it is possible to extract about $\rho_\infty$ bits per symbol of information by observing the sequences $\{X_k\}$ and $\{Y_k\}$ independently. Gács and Körner [13] defined a notion of CI as $C_{GK}(X;Y) = \sup \rho_\infty$, where the supremum is taken over all sequences of pairs of mappings for which $\lim_{n\to\infty} \varepsilon_n = 0$. Gács and Körner [13] and later Witsenhausen [14] showed that the common core of two correlated discrete memoryless information sources (2-DMS) is not always "materializable", even if the sequences $\{X_i\}_{i=1}^{\infty}$ and $\{Y_i\}_{i=1}^{\infty}$ are highly correlated, i.e., $C_{GK}(X;Y) = 0$, except for the special case where $X = (U, Q)$ and $Y = (V, Q)$, and $H(Q) > 0$. Remarkably, they showed that the asymptotic case is no better than the zero-error case and $C_{GK}(X;Y) = H(\text{mcf}(X,Y)) = H(X \wedge Y)$. Thus, common codes of a 2-DMS cannot exploit any correlation beyond a certain deterministic interdependence of the sources. $X$ and $Y$ can be highly correlated and yet have nothing *explicitly in common*.

The zero pattern of the joint distribution $p_{XY}$ can be specified by its characteristic bipartite graph $B_{XY}$ [19], [20], [21] with the vertex set $\mathcal{X} \cup \mathcal{Y}$ and an edge connecting two vertices $x$ and $y$ if $p_{XY}(x,y) \geq 0$. Following the terminology in [21], suppose that the graph thus obtained contains $k$ connected components (called *minimal disjoint components*, MDCs), where $1 \leq k \leq |\mathcal{X}|$. Now let the function $f: \mathcal{X} \to 2^{\mathcal{X} \cup \mathcal{Y}}$ map a vertex $m \in \mathcal{X}$ of $B_{XY}$ into the set of vertices in the MDC of $B_{XY}$ containing $m$. Similarly, define the function $g: \mathcal{Y} \to 2^{\mathcal{X} \cup \mathcal{Y}}$. The common RV of $X$ and $Y$ is then $Q = X \wedge Y = f(X) = g(Y)$. Now, every pair $(X,Y)$ admits a unique decomposition of $(\mathcal{X}, \mathcal{Y})$ into MDCs. Let $\{(\mathcal{X}_1, \mathcal{Y}_1), \ldots, (\mathcal{X}_k, \mathcal{Y}_k)\}$ be such a decomposition, where $\mathcal{X}_i \subseteq \mathcal{X}, \mathcal{Y}_i \subseteq \mathcal{Y}$. Defining $p_Y(\mathcal{Y}_i) \triangleq \sum_{y \in \mathcal{Y}_i} p_Y(y)$, the Gács and Körner CI can alternatively be defined as $H(X \wedge Y) = \sum_{i=1}^{k} H(p_Y(\mathcal{Y}_i))$ (Theorem 5, [21]). It is easy to check that $H(X \wedge Y) = 0$ iff $(\mathcal{X}, \mathcal{Y})$ is a MDC (i.e., $k = 1$), when we say that $p_{XY}$ is *indecomposable*. For example, a binary symmetric channel with non-zero crossover probability has a single MDC, and hence $H(X \wedge Y) = 0$. Now define the zero information component (ZIC) as follows: $(\mathcal{X}_i, \mathcal{Y}_i)$ is a ZIC if $\forall x \in \mathcal{X}_i, y, y' \in \mathcal{Y}_i$, $p_{X|Y}(x|y) = p_{X|Y}(x|y')$, and $\forall x \notin \mathcal{X}_i, y \in \mathcal{Y}_i, p_{Y|X}(y|x) = 0$. $H(X \wedge Y) = I(X;Y)$, iff all MDCs are ZICs [21], when we say that the pair $(X,Y)$ is *perfectly resolvable* [16]. We say that the common core $Q$ perfectly resolves $(X, Y)$, if $I(X;Y|Q) = 0$ and $H(Q|X) = H(Q|Y) = 0$. Figure 2 shows the bipartite graph of a pair of dependent RVs $(X,Y)$. The solid black lines each have a probability mass of $\frac{1-\delta}{8}$, and the lighter ones $\frac{\delta}{8}$. When $\delta = 0$, $X$ and $Y$ can be written as $(U,Q)$ and $(Q,V)$ respectively, where $U$, $V$, $Q$ are independent, and $C_{GK}(X;Y) = H(Q) = 1$. Intuitively, for small values of $\delta$, $(X,Y)$ have still a lot in common. However $C_{GK}(X;Y)$ is discontinuous in $\delta$, jumping from $C_{GK}(X;Y) = 1$ at $\delta = 0$ to $C_{GK}(X;Y) = 0$ for any $\delta > 0$, howsoever small [13], [14], [16]. Thus, when there is only a single connected component, $C_{GK}(X;Y) = 0$ even if it is the case that by removing a set of edges that account for a small probability mass, the graph can be decomposed into a large number of MDCs, each with a significant probability mass.

**Figure 2.** Bipartite graph of $(X,Y)$.

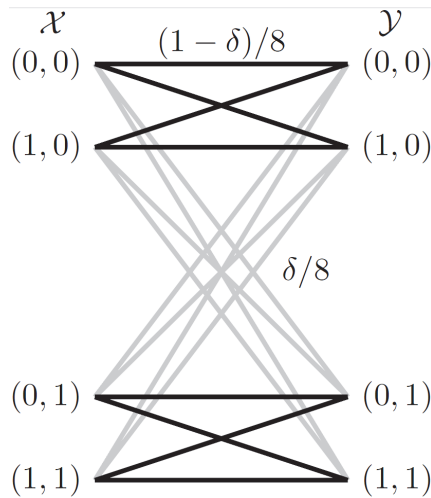

### 2.3. Wyner common information and more recent generalizations

The fact that $C_{GK}(X;Y)$ is degenerate for most non-trivial distributions motivated Wyner [15] to devise a distributed lossless source coding setup for coding the 2-DMS into a common part (analogous to $Q$) and two "private" parts (analogous to $U$ and $V$). He gave a single-letter characterization of the CI that is strictly greater than mutual information as follows: $C_W(X;Y)$ is the infimum of the mutual information between $(X, Y)$ and $Q$, where the infimum is taken over all auxiliary RVs $Q$ conditioned on which $X$ and $Y$ are independent, i.e., $C_W(X;Y) = \inf_{Q: X-Q-Y} I(XY;Q)$, with $|\mathcal{Q}| \leq |\mathcal{X}||\mathcal{Y}|$, where $X - Q - Y$ means that these RVs form a Markov chain in that order. $C_W(X;Y)$ quantifies the resource cost of generating correlated RVs.

The different notions of CI are related as, $C_{GK}(X;Y) \leq I(X;Y) \leq C_W(X;Y)$, with equality holding iff $(X,Y)$ can be written in the form $X = (U,Q)$ and $Y = (V,Q)$ where $U, V, Q$ are independent, whence $C_{GK}(X;Y) = I(X;Y) \Leftrightarrow I(X;Y) = C_W(X;Y)$. While the Gács and Körner CI obeys an intuitive monotonically non-increasing property that any reasonable definition of common information must satisfy, Wyner's CI is a non-decreasing function of the number of input arguments [22]. In particular, the following monotonicity relationships are well-known [22], [23].

$$\begin{aligned}
&C_{GK}(X_1;\ldots;X_K) \leq C_{GK}(X_1;\ldots;X_{K-1}) \\
&C_W(X_1;\ldots;X_K) \geq C_W(X_1;\ldots;X_{K-1}) \\
&C_{GK}(X_1;\ldots;X_K) \leq \min_{i \neq j,\, i,j=1,\ldots,K} I(X_i;X_j) \leq \max_{i \neq j,\, i,j=1,\ldots,K} I(X_i;X_j) \leq C_W(X_1;\ldots;X_K)
\end{aligned} \quad (1)$$

Very recently, a generalization of the setup of Gács and Körner was developed in [16] where a three-dimensional rate region called the *assisted residual information region*, $\mathfrak{I}(X;Y)$ is introduced.

The rate region quantifies the extent to which a piece of common information resolves the dependence between $(X,Y)$, relying on rate-limited private communication from an omniscient genie to unlock hidden layers of "almost common" information. The boundary of the $\mathfrak{T}(X;Y)$ region is made up of triples of the form $(\Delta_{21},\Delta_{22},\Delta_1) = (I(Y;Q|X), I(X;Q|Y), I(X;Y|Q))$. If the $\mathfrak{T}(X;Y)$ region includes the origin, then the common core $Q$ perfectly resolves $(X, Y)$, whence $C_{GK}(X;Y) = I(X;Y)$, i.e., conditioned on $Q$, there is no "residual information" that correlates $X$ and $Y$. Both the residual information $(I(X;Y) - C_{GK}(X;Y))$ and Wyner CI feature as extreme points on the boundary of the $\mathfrak{T}(X;Y)$ region [16] and can be expressed as follows:

$$\begin{aligned} I(X;Y) - C_{GK}(X;Y) &= \min_{Q:\, H(Q|X)=H(Q|Y)=0} I(X;Y|Q) = \min_{(0,0,R_0) \in \mathfrak{T}(X;Y)} R_0 \\ C_W(X;Y) &= I(X;Y) + \min_{Q:\, X-Q-Y} (I(Y;Q|X) + I(X;Q|Y)) = I(X;Y) + \min_{(R_1,R_2,0) \in \mathfrak{T}(X;Y)} R_1 + R_2 \end{aligned} \qquad (2)$$

Importantly, the nontrivial shape of the boundary of the $\mathfrak{T}(X;Y)$ region captures the subtle characteristics of correlation that is not reflected in the common information of Gács and Körner.

## 3. Quantifying redundant information

Let $X_{\mathcal{A}} \triangleq \{X_a\}_{a \in \mathcal{A}}$ be a $K$-tuple of RVs ranging over finite sets $\mathcal{X}_a$, where $\mathcal{A}$ is an index set of size $|\mathcal{A}| = K$. Now let $X_{\mathcal{A}_i} = \{X_a\}_{a \in \mathcal{A}_i}$ be $m$ non-empty, potentially overlapping subsets of $X_{\mathcal{A}}$ with index sets $\mathcal{A}_i \subseteq \mathcal{A}$, and define $X_{\mathcal{R}} \triangleq \{X_{\mathcal{A}_i}\}_{\mathcal{A}_i \subseteq \mathcal{A}}$ as any arbitrarily chosen collection of $m$ such subsets.

Williams and Beer [1] introduced the partial information (PI) diagrams to decompose the total mutual information, $I(X_{\mathcal{R}};Y)$ into non-negative, non-overlapping partial information (PI) atoms s.t. summing over all the PI-atoms yields back the total mutual information $I(X_{\mathcal{R}};Y)$. Figure 3 shows the PI-diagram for $K = 2$ [1], [11]. Each irreducible PI-atom represents information that is redundant, unique or synergistic. Following the notation introduced in [11], e.g., for $K = 2$, {1} and {2} are respectively, the unique information about $Y$, that $X_1$ and $X_2$ exclusively convey; {1, 2} is the redundant information about $Y$, that $X_1$ and $X_2$ both convey; {12} is the synergistic information about $Y$, that the joint RV $(X_1, X_2)$ conveys, i.e., information that can only be conveyed by a coalition.

From Figure 3, it is easy to see that the three equations specifying $I(X_1 X_2;Y)$, $I(X_1;Y)$ and $I(X_2;Y)$ do not fully determine the four functions {1,2}, {1}, {2} and {12}. Williams and Beer [1] accomplish the decomposition by defining a notion of redundant information and proposed the following axioms that any valid measure of redundancy $I_{\cap}(X_{\mathcal{R}};Y)$ must satisfy:

1. **(GP)** *Global positivity*: $I_{\cap}(X_{\mathcal{R}};Y) \geq 0$.
2. **(S)** *Weak symmetry*: $I_{\cap}(X_{\mathcal{R}};Y)$ is symmetric in $X_{\mathcal{R}}$.
3. **(I)** *Self-redundancy*: $I_{\cap}(X_1;Y) = I(X;Y)$.
4. **(M)** *Monotonicity*: $I_{\cap}(X_{\mathcal{A}_1},\ldots,X_{\mathcal{A}_m};Y) \leq I_{\cap}(X_{\mathcal{A}_1},\ldots,X_{\mathcal{A}_{m-1}};Y)$.

Following [1], a host of other desirable axioms have been proposed [2], [24], [10]. However, for the exposition to follow, it suffices to consider the above properties. While defining the collection of subsets $X_{\mathcal{R}}$, we did not rule out the possibility that some subsets might be supersets of others. Properties **(S)** and **(M)** enormously simplify the bookkeeping structure in that only those subsets need to be considered which satisfy the ordering relation $\mathcal{A}_i \not\subseteq \mathcal{A}_j, \forall i \neq j$ [1], [2]. The redundancy structure then naturally induces a partial order on all such valid subsets $X_{\mathcal{R}_v} \subseteq X_{\mathcal{R}}$, so that $I_\cap(X_{\mathcal{R}_v};Y)$ is now a monotone function on the lattice of anti-chains [1]. Williams and Beer [1] called such a lattice structure of redundancy as the *partial information lattice*. Thus when $I_\cap(.)$ is defined, a unique decomposition can be accomplished using a Möbius inversion [1], taking care to ensure that locally all the PI-atoms are non-negative (local positivity).

**Figure 3.** PI-diagram for $K = 2$.

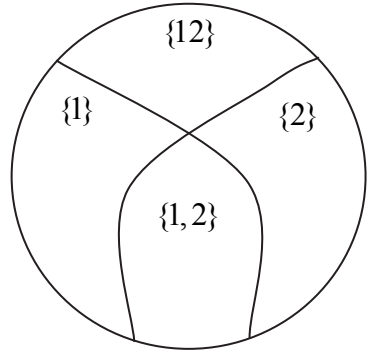

$$I(X_1X_2;Y) = \underbrace{\{1,2\}}_{redundant} + \underbrace{\{1\}+\{2\}}_{unique} + \underbrace{\{12\}}_{synergistic}$$

$$I(X_1;Y) = \{1,2\} + \{1\}$$
$$I(X_2;Y) = \{1,2\} + \{2\}$$
$$I(X_1;Y \mid X_2) = \{1\} + \{12\}$$
$$I(X_2;Y \mid X_1) = \{2\} + \{12\}$$

As elegant a decomposition as it is, the PI decomposition is however not perfect. Increasing the number of predictor RVs amounts to a combinatorial explosion of PI-atoms. Already in the $K = 3$ case, there are 18 PI-atoms, since the same state of the target RV can now have any combination of redundant, unique or synergistic PI-atoms [11]. For instance, now the PI-atoms comprise of the following combinations:
- unique: $\{1\},\{2\},\{3\}$;
- redundant: $\{1,2\},\{2,3\},\{3,1\},\{1,2,3\},\{1,23\},\{2,13\},\{3,12\}$; and
- synergistic: $\{12\},\{23\},\{31\},\{123\},\{12,13\},\{12,23\},\{13,23\},\{12,23,31\}$.

Attributing operational meaning to each of these atoms is a significant challenge. By considering a simple three-input XOR example, it was shown in [2], that local positivity is incompatible with the PI lattice. Furthermore, the redundancy measure proposed by Williams and Beer [1] suffered from some important drawbacks [2], [24]. Since then, several other notions of redundancy have been proposed. Using the framework of information geometry, Harder *et al.* [24] defined a measure of bivariate redundancy based on projections in the space of probability distributions.

More recently, Griffith *et al.* [10] defined a measure of redundancy using the Gács and Körner's CI. They defined the zero-error redundant information shared between the $K$-tuple of predictor RVs $X_\mathcal{A} = \{X_a\}_{a \in \mathcal{A}}$ and the target RV $Y$, $C_\cap^{GK}(X_\mathcal{A};Y)$ as

$$C_{\cap}^{GK}(X_{\mathcal{A}};Y) \triangleq \max_{Q: H(Q|X_a)=0, \forall a \in \mathcal{A}} C_{GK}(Q;Y)$$

$$\overset{(a)}{=} C_{GK}(X_1 \wedge ... \wedge X_K;Y) \quad (3)$$

$$= H(X_1 \wedge ... \wedge X_K \wedge Y)$$

where (a) follows from the fact, $Q = (X_1 \wedge ... \wedge X_K)$ is the largest common RV that is informationally poorer than $X_a \; \forall a \in \mathcal{A}$, i.e., $Q \preccurlyeq X_a \; \forall a \in \mathcal{A}$ (in the sense of the partial order $\preccurlyeq$). Based on the above definition, they defined the mutual information counterpart $I_{\cap}^{GK}(X_{\mathcal{A}};Y)$ as

$$I_{\cap}^{GK}(X_{\mathcal{A}};Y) \triangleq I(X_1 \wedge ... \wedge X_K;Y) \quad (4)$$

Griffith *et al.* [10] showed that both $C_{\cap}^{GK}(X_{\mathcal{A}};Y)$ and $I_{\cap}^{GK}(X_{\mathcal{A}};Y)$ satisfy a host of important properties (see Table 1 in [10]) associated with an ideal redundancy measure. Though elegant in its formulation, $I_{\cap}^{GK}(X_{\mathcal{A}};Y)$ inherits the negative character of the original definition of Gács and Körner and is trivially zero for a large class of interesting joint distributions. Thus, unless the *K*-tuple of predictors have a *decomposable* joint distribution, $I_{\cap}^{GK}(X_{\mathcal{A}};Y)$ is zero, even if it is the case that the predictors share non-trivial redundant information about the target *Y*.

It is instructive to consider if Wyner's CI is a better alternative in specifying redundant information. Unfortunately, as is clear from (1), Wyner's definition violates the monotonicity property **(M)** and is not a suitable measure of redundancy in the PI decomposition framework. Perhaps, a more useful motion of common information is captured in the assisted residual information setup [16]. For triples of the form $(I(Y;Q|X), I(X;Q|Y), I(X;Y|Q))$ on the boundary of the $\mathfrak{T}(X;Y)$ region, define the *minimum assisted common information* as the minimum distance (in a Euclidean distance sense) from the origin to the boundary of the $\mathfrak{T}(X;Y)$ region as follows:

$$C_{\min}(X;Y) \triangleq \inf_Q (I(Y;Q|X) + I(X;Q|Y) + I(X;Y|Q)) \quad (5)$$

$C_{\min}(X;Y)$ measures the minimum distance from perfect resolvability and captures the notion of minimal assistance required from a genie so that the common core can be extracted for all non-trivial correlations. However, unlike $C_{GK}(X;Y)$, $C_{\min}(X;Y)$ appears to be a much more difficult quantity to compute for interesting RVs. Moreover, $C_{\min}(X;Y)$ violates the monotonicity property **(M)** and for all practical purposes is restricted to the bivariate case.

## 4. Quantifying unique information

Let $\{X_i, Y_i\}_{i=1}^{\infty}$ be i.i.d. copies of the pair $(X, Y) \sim p(x, y)$, taking values from finite sets $\mathcal{X}$ and $\mathcal{Y}$. To transmit the sequence $\{X_i\}$ via a finite capacity channel, with asymptotically negligible probability of error, the channel capacity must at least be $H(X)$. However, if side-information $\{Y_i\}$ is made available at the decoder, then the celebrated Slepian-Wolf theorem ([25], Theorem 15.4.1) says that the required capacity is reduced to $H(X|Y)$. Thus it is tempting to interpret $H(X|Y)$ as a natural measure of unique or private information of *X* with respect to *Y*, that can be isolated and transmitted separately. Completely describing *X* thus entails no more than first describing the information that *Y* shares about

$X$ at a rate $R_Y = I(X;Y)$ and then describing the remaining uncertainty about $X$ at rate $H(X|Y)$. However, contrary to this seemingly harmless intuition, explicit examples were constructed in [21], where $H(Y)$ is arbitrarily large, $Y$ contains arbitrarily small information about $X$ and yet extracting this infinitesimally small amount of information, $I(X;Y)$, requires a complete description of $Y$, i.e., $R_Y \geq H(Y) \gg I(X;Y)$. This implies that the information contained in $Y$ about $X$ cannot be separated from the other information that $Y$ contains [21], and hence unique information cannot be extracted in general. Given such subtleties involved with its asymptotic counterparts, it is natural to ask if there exists a zero-error version for private information of $X$ with respect to $Y$.

*4.1. A combinatorial dual of Gács and Körner's common information*

In the late 1970s, Witsenhausen [19] explored a zero-error side information problem for correlated sources. He showed that, in the presence of side information $Y$ at the decoder, the minimum cardinality of the signal alphabet needed to transmit $X$ *without any error* is related to the chromatic number of the characteristic graph $G_{XY} = (\mathcal{X}, \mathcal{E})$. $G_{XY}$ is derived from the bipartite graph $B_{XY}$ introduced earlier as follows: for distinct vertices $x_1, x_2 \in \mathcal{X}$, $(x_1, x_2) \in \mathcal{E}$, iff $\exists y \in \mathcal{Y}$, s.t. $p_{XY}(x_1, y) > 0$, $p_{XY}(x_2, y) > 0$, we call such a pair of vertices $(x_1, x_2) \in \mathcal{E}$ as *confusable* [19]. Then the chromatic number of $G_{XY}$ gives the minimal number of symbols needed to transmit $X$ and the minimal such valid coloring gives the private information of $X$ with respect to $Y$, $PI_W(X \setminus Y)$, i.e., amount of information that $X$ needs to transmit to $Y$ in order to share information with zero error. $PI_W(X \setminus Y)$ may be construed as the minimal amount of information that $Y$ needs from $X$ to fully describe $X \vee Y$, i.e., $PI_W(X \setminus Y) \vee Y = X \vee Y$.

The following contrived example illustrates the concepts: Consider the set, $\Omega = \{0,1,2,3,4,5,6,7,8,9,a,b,c,d,e,f\}$. Since information elements can be identified with sample-space partitions, consider now the following partitions on the set $\Omega$:

$$X = 01\,|\,23\,|\,45\,|\,67\,|\,89\,|\,ab\,|\,cd\,|\,ef, \text{ and } Y = 02\,|\,34\,|\,56\,|\,78\,|\,19\,|\,ac\,|\,de\,|\,bf$$

It is easy to see that

$$X \vee Y = 0\,|\,1\,|\,2\,|\,3\,|\,4\,|\,5\,|\,6\,|\,7\,|\,8\,|\,9\,|\,a\,|\,b\,|\,c\,|\,d\,|\,e\,|\,f, \text{ and}$$
$$X \wedge Y = 0123456789\,|\,abcdef$$

Figure 4 shows the bipartite graph $B_{XY}$ consisting of two disjoint noisy typewriter channels and the characteristic graph $G_{XY}$ derived from it. The confusable vertex pairs include

$$(01, 23), (23, 45), (45, 67), (67, 89), (89, 01), (ab, cd), (cd, ef), \text{ and } (ab, ef).$$

Several distinct valid minimal colorings exist [19], two of which are shown below, along with the induced *σ*-algebras:

$$\gamma_1 = \{01, 67, ab\}, \ \gamma_2 = \{23, 89, ef\}, \ \gamma_3 = \{45, cd\}$$
$$PI_W(X \setminus Y) = 0167ab\,|\,2389ef\,|\,45cd$$

$$\gamma_1' = \{01, 45, ab\}, \ \gamma_2' = \{23, 67, cd\}, \ \gamma_3' = \{89, ef\}$$
$$PI_W'(X \setminus Y) = 0145ab\,|\,2367ef\,|\,89cd$$

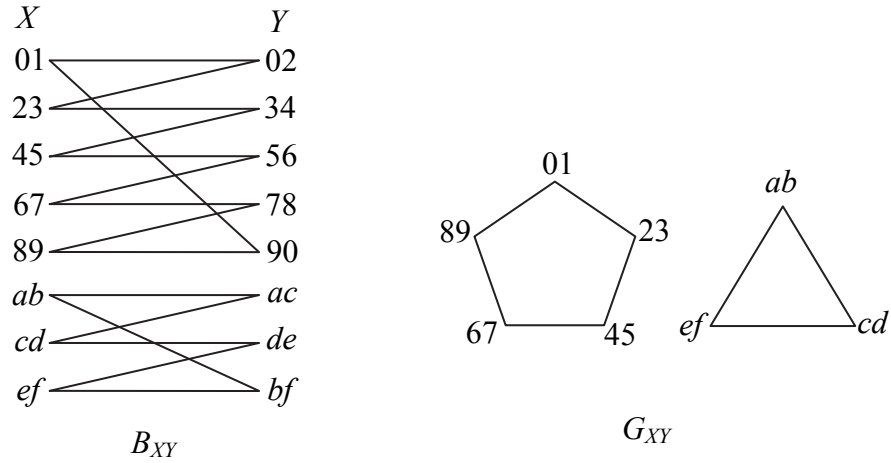

**Figure 4.** The bipartite graph $B_{XY}$ consisting of two disjoint noisy typewriter channels and the characteristic graph $G_{XY}$ derived from it.

It is easy to check that

$$PI_W(X \setminus Y) \vee Y = PI'_W(X \setminus Y) \vee Y = X \vee Y$$

Hence, such a minimal coloring is not unique [19] and consequently, $PI_W(X \setminus Y)$ is not unique.

Hexner and Yo [26], [27] studied common and private information structures in a decision-theoretic framework. Their definition of common information coincides with that of Gács and Körner's definition [13]. However, their notion of private information differs from that of Witsenhausen [19]. They defined private information, $PI_{HY}(X \setminus Y)$ as the minimal amount of information needed to fully describe $X$, given the common information $X \wedge Y$, i.e., $PI_{HY}(X \setminus Y) \vee (X \wedge Y) = X$. Unfortunately, the definition does not admit a unique specification for the private information as can be seen from the following example [26]. Consider the set, $\Omega = \{0,1,2,3,4,5\}$, and the following partitions on the set $\Omega$.

$$X = 0|12|3|45, \text{ and } Y = 01|2|34|5$$

It is easy to see that,

$$X \vee Y = 0|1|2|3|4|5, \text{ and}$$
$$X \wedge Y = 012|345$$

Further, each of the following $\sigma$-algebras satisfies the definition, i.e.

$$PI_{HY}(X \setminus Y) = 03|1245$$
$$PI'_{HY}(X \setminus Y) = 045|123$$

$$PI_{HY}(X \setminus Y) \vee (X \wedge Y) = PI'_{HY}(X \setminus Y) \vee (X \wedge Y) = X$$
$$PI_{HY}(X \setminus Y) \vee Y = PI'_{HY}(X \setminus Y) \vee Y = X \vee Y$$

Thus, like $PI_W(X \setminus Y)$, $PI_{HY}(X \setminus Y)$ is not unique. The two definitions may be contrasted in that, while $PI_{HY}(X \setminus Y)$ complements $X \wedge Y$ to reconstruct $X$, $PI_W(X \setminus Y)$ complements $Y$ to reconstruct $X \vee Y$. Hexner and Yo [27] surmised that much of the pathological behavior of the private information structures can be attributed to the non-modularity of the information lattice. In what follows, we analyze some of the more recent attempts to quantify unique information, mostly pursued in the context of the PI decomposition framework.

*4.2. The union of parts and the intrinsic conditional information*

Griffith and Koch [11] noted that by simply amending the colloquial definition of synergy from "whole minus the *sum* of parts" to "whole minus the *union* of parts" one can discount double-counting whenever there is redundancy among the parts. Using the two-input XOR as an exemplary model of synergy, they adapted the intrinsic conditional information [28] to compute the synergy as a constrained optimization problem. To see this, notice that for a two-input XOR, $Y = \text{XOR}(X_1, X_2)$, where $X_i \sim \text{Bernoulli}(\frac{1}{2})$, $i = 1, 2$, the whole of $(X_1 X_2)$ is required to fully specify $Y$, i.e., $I(X_1; X_2) = I(X_1; Y) = I(X_2; Y) = 0, I(X_1 X_2; Y) = H(Y) = 1$. Now consider the following information-theoretic identity, where we have expanded the terms on the right using the PI-diagram in Figure 3.

$$I(X_1; X_2 | Y) = \underbrace{I(X_1 X_2; Y)}_{\{12\}+\{1\}+\{2\}+\{1,2\}} + \underbrace{I(X_1; X_2)}_{\{1,2\}} - \underbrace{I(X_1; Y)}_{\{1\}+\{1,2\}} - \underbrace{I(X_2; Y)}_{\{2\}+\{1,2\}} = \{12\}$$

It appears that the conditional mutual information $I(X_1; X_2 | Y)$ is an ideal candidate measure to quantify synergy, $\{12\}$. However, it is well known that whereas conditioning always reduces ordinary entropy, the same does not hold for the mutual information, unless the triple of RVs $(X_1, X_2, Y)$ form a Markov chain in any order [25]. In fact, whenever, we have pairwise independence, conditioning always *increases* mutual information. This is easily seen for a two-input XOR example since $I(X_1; X_2) = 0$, whereas $I(X_1; X_2 | Y) = 1$. Thus, conditioning on the side information $Y$ generates *artificial* correlation beyond that is actually there. Maurer and Wolf [28] utilize this simple argument to construct their upper bound on the secret key rate by taking a reduced conditional mutual information, where the reduction is effected over all Markov chains $X_1 X_2 - Y - Y'$, i.e.,

$$I(X_1; X_2 \downarrow Y) = \min_{X_1 X_2 - Y - Y'} I(X_1; X_2 | Y'),$$

where the cardinality of the alphabet $\mathcal{Y}'$ of RV $Y'$ is bounded as $|\mathcal{Y}'| \leq |\mathcal{Y}|$. The newly derived quantity, called the *intrinsic conditional information*, $I(X_1; X_2 \downarrow Y)$ discounts all artificial correlations brought in by the additional side-information $Y$. Indeed for the XOR example, $I(X_1; X_2 | Y) = 1$, but $I(X_1; X_2 \downarrow Y) = 0$. Intuitively, these artificial correlations are the source of synergy as realized by Griffith and Koch who define synergy as $I_{Synergy}(X_1 X_2; Y) = (I(X_1; X_2 | Y) - I(X_1; X_2 \downarrow Y))$ to quantify the whole minus the union of parts. In what follows, we recount their arguments showing that the constrained optimization set they arrive at leads to exactly the same definition of unique information as proposed by Bertschinger *et al.* [12]. Finally, we mention at least one scenario, where their measure can fail.

In the derivation to follow, we utilize the decomposition of $I(X_1; X_2 | Y)$ into PI-atoms (see Figure 3).

$$I(X_1; X_2 | Y) - I(X_1; X_2 \downarrow Y)$$
$$= I(X_1; X_2 | Y) - \min_{X_1 X_2 - Y - Y'} I(X_1; X_2 | Y')$$
$$= I(X_1; X_2 | Y) - \min_{X_1 X_2 - Y - Y'} [I(X_1 X_2; Y') + I(X_1; X_2) - I(X_1; Y') - I(X_2; Y')]$$
$$\stackrel{(a)}{=} I(X_1 X_2 | Y) - I(X_1; Y) - I(X_2; Y) - \min_{X_1 X_2 - Y - Y'} \Big[ \underbrace{I(X_1 X_2; Y') - I(X_1; Y') - I(X_2; Y')}_{\{12\} - \{1,2\}} \Big]$$
$$\stackrel{(b)}{=} I(X_1 X_2 | Y) - I(X_1; Y) - I(X_2; Y) - \min_{\substack{X_1 X_2 - Y - Y' \\ \Pr\{X_i, Y\} = \Pr\{X_i, Y'\}, i=1,2}} \Big[ \underbrace{I(X_1 X_2; Y') - I(X_1; Y') - I(X_2; Y')}_{\{12\} - \{1,2\}} \Big]$$
$$= I(X_1 X_2 | Y) - \min_{\substack{X_1 X_2 - Y - Y' \\ \Pr\{X_i, Y\} = \Pr\{X_i, Y'\}, i=1,2}} I(X_1 X_2; Y')$$
$$= I_{Synergy}(X_1 X_2; Y)$$

Note that, in going from steps (a) to (b), additional constraints have been imposed to ascribe operational meaning to the optimal auxiliary RV $Y'$ such that $Y'$ specifies the union of parts. One way to do this is by noting that both the unique and redundant information, and hence the union of parts information should depend only on the marginal distribution of the pairs $(X_1, Y)$ and $(X_2, Y)$ [11], [12]. Preserving the consistency of the marginal distributions then amounts to the following constraints: $\Pr\{X_i, Y\} = \Pr\{X_i, Y'\}, i = 1, 2$. Also, these additional constraints do not increase the argument under $\min_{X_1 X_2 - Y - Y'}[.]$, i.e., $\{12\} - \{1, 2\}$, which is the difference between the synergy and the redundancy PI-atoms [11]. It may be noted that Bertschinger *et al.* [12] approach the same problem of quantifying unique and redundant information utilizing a decision-theoretic framework. In particular, they use a similar set of constraints in motivating an operational meaning of unique and redundant information. The remainder of the derivation follows, given these additional constraints.

Griffith and Koch's measure, $I_{Synergy}(X_1 X_2; Y)$ suffers from an important drawback in that, it is *lockable*, i.e., bringing in an additional target RV $Y_1$ can increase the synergy sharply by an arbitrarily large amount. This can be seen from the fact that, while additional side information cannot reduce the conditional mutual information by more than the entropy of the side information, i.e., $|I(X_1; X_2 | Y Y_1) - I(X_1; X_2 | Y)| \leq H(Y_1)$, the same does not hold for the intrinsic conditional information [32]. Thus, an extra conditioning on $Y_1$ can increase the difference $(I(X_1; X_2 | Y Y_1) - I(X_1; X_2 \downarrow Y Y_1))$ sharply, which is clearly not a desirable property of an ideal measure of synergy. Possibly, considering multiple adaptive local minimizations of the mutual information [32] instead of a single global one can remedy this concern.

*4.3. Conditional Gács and Körner's CI and unique information*

It may be noted that under a reordering of the arguments, the intrinsic information, $I(Y; X_1 \downarrow X_2)$ has a more natural analogy with the notion of unique information, since this quantity naturally quantifies how much information that $X_1$ has about $Y$ that is unknown to $X_2$. However, as can be easily checked, $I(Y;X_1 \downarrow X_2)$ does not obey an important consistency condition, that any valid measure of unique information $I_{unique}$ must satisfy [2], viz.,

$$I(Y;X_1) + I_{unique}(Y;X_2 | X_1) = I(Y;X_2) + I_{unique}(Y;X_1 | X_2) \tag{6}$$

Hence, $I(Y;X_1 \downarrow X_2)$ cannot be interpreted as unique information in the PI decomposition framework.

A conditional version of the Gács and Körner's CI is defined as $C_{GK}(Y;X_1 | X_2)$ $\triangleq \max_{Q: Q-YX_2-X_1; Y-X_1X_2-Q} I(YX_1;Q | X_2)$. $C_{GK}(Y;X_1 | X_2)$ satisfies the aforementioned consistency condition in (6) when some additional constraints are imposed on the underlying distribution. This is best illustrated with an example. Suppose $X_1, X_2$ are two binary RVs chosen independently and uniformly from $\{0,1\}$, and $Y$ is an unaltered copy of both $X_1$ and $X_2$. Clearly, both $X_1$ and $X_2$ contain 1 bit worth of unique information about $Y$, whereas there is no redundancy between $X_1$ and $X_2$ with respect to $Y$, since $X_1$ and $X_2$ are independent. This intuition is borne out of the fact that $C_{GK}(Y;X_1 | X_2) = C_{GK}(Y;X_2 | X_1) = 1$ bit. More generally, it is easy to check that if the pairs $(Y, X_1)$ and $(Y, X_2)$ are perfectly resolvable, then $C_{GK}(Y;X_1 | X_2)$ and $C_{GK}(Y;X_2 | X_1)$ satisfies the consistency condition

$$I(Y;X_1) + C_{GK}(Y;X_2 | X_1) = I(Y;X_2) + C_{GK}(Y;X_1 | X_2) \tag{7}$$

and hence is an ideal measure of unique information, albeit under highly restrictive assumptions.

*4.4. Information bottleneck, sufficient statistics and unique information*

In Section 4.2, we have seen that in quantifying unique information, both the approaches in [11] and [12] involve some kind of a constrained optimization. In this section, we seek to answer the following question: what other constrained optimization setups can be used to quantify unique information? It turns out that, under some additional constraints on the underlying distribution, we can quantify unique information using a modified framework of the information bottleneck (IB) principle [29], called the information bottleneck with side information (IBSI) [30]. In particular, for the two-variable case, by treating the information provided by one singleton predictor as *irrelevant* side information, unique information provided by the other singleton can be extracted as *relevant* information. In [31], an improved formulation called the conditional information bottleneck (CIB) was introduced. Recall that for any Markov chain of the form, $P-X-f(X)$, the data processing inequality [25] guarantees that $I(P;X) \geq I(P;f(X))$. In case equality holds, i.e., if $P-f(X)-X$, we say that $f(X)$ is an *exactly* sufficient statistic for predicting $P$ from $X$. Finding an *exactly* sufficient statistic is a difficult problem, but the IB framework provides a tractable alternative to finding *approximately* sufficient statistic [29].

Before getting into the details of extraction of unique information using the modified IB framework, we discuss an important practical limitation of estimating the joint distribution of the set of all predictors and the target. Sample sizes, for instance, from typical recordings in electrophysiological experiments are rarely sufficient for reliable estimation of the entire joint distribution even for moderate values of $K$. It turns out that the assumption of target stimulus-conditioned independence of the predictors simplifies the picture to a considerable extent without deviating much from the true joint distribution [30]. Such an assumption is intrinsic to both the IBSI and CIB problem setups and with this additional constraint on the underlying joint distribution, unique information can be extracted by either formulation. The discussion below follows the CIB formulation.

The CIB problem [31] is formulated as follows: Given the joint distribution $p(x_1, x_2, y)$ of the pair $(X_1, X_2, Y)$, suppose the goal is to extract the unique information that singleton predictor $X_1$ provides about $Y$. This can be accomplished by finding a stochastic map $p(t|y)$ from $Y$ to a "bottleneck" RV $T$, s.t., $T$ maximizes the conditional mutual information $I(T; X_1 | X_2)$. $T$ only depends on $Y$ and is independent of $X_1$ and $X_2$, given $Y$. Here $X_1$ is called the *relevant* variable and $X_2$ the *irrelevant* side information. In effect, the unique information that $X_1$ provides about $Y$ is *squeezed* through the compact bottleneck representation $T$. This idea is illustrated in Figure 5. It may be noted that the above Markov chain constraint is not a modeling assumption, but rather a part of the definition of the problem setup, so that the marginal over $p(x_1, x_2, y, t)$ is always consistent with the input distribution $p(x_1, x_2, y)$ [29]. This then leads to a constrained optimization problem that can be formulated using Lagrange multipliers as follows [31],

$$\min_{p(t|y)} \mathcal{L} \triangleq I(Y;T) - \beta I(X_1; T | X_2), \text{ s.t. } \sum_t p(t|y) = 1, \forall y \text{ and } p(t|y) \geq 0, \forall y, t \quad (8)$$

The parameter $\beta$ captures the level of quantization we can afford to tolerate in approximating an exactly sufficient statistic. As $\beta \to 0$, we are mostly interested in maximal compression, so that everything is assigned to a single point, and $I(T;Y) = 0$. As $\beta \to \infty$, arbitrarily detailed quantization can be achieved. For any $\beta$ in between these two extremes, we can explore the trade-off between preserving the relevant unique information (that $X_1$ provides about $Y$) and compression of $Y$ at various resolutions. While the underlying optimization problem is not convex, convergence is still guaranteed owing to the elegance of the IB variational formulation [31], [29]. The solution is characterized by a set of self-consistent equations that yields an iterative algorithm that is guaranteed to converge (at least locally) by alternating iterations amongst a set of convex distributions [31], [29].

**Figure 5.** Extracting the unique information that singleton predictor $X_1$ (relevant variable) provides about $Y$ (given side information in the form the irrelevant variable $X_2$) using the conditional information bottleneck principle. In effect, the unique information that $X_1$ provides about $Y$ is *squeezed* through the compact bottleneck representation $T$

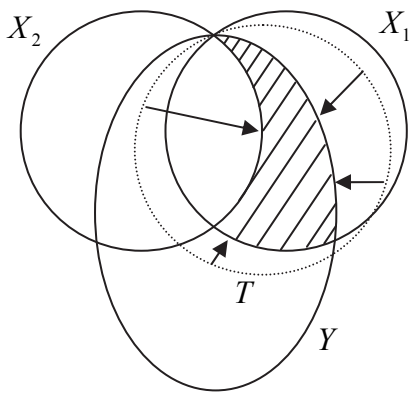

## 5. Conclusions

In this paper, we took a closer look at the structure of bivariate dependency induced by a pair of predictor RVs trying to encode a target RV *Y*. It remains unclear, whether a desired decomposition in the PI framework [1] should be based on Gács and Körner's notion of CI [1]. While the latter notion enjoys the unique property of being representable as an information partition [17], [9], only for a special class of decomposable distributions such a measure yields useful results. A related measure, Wyner's CI is a non-decreasing function of the number of input arguments and does not satisfy an intuitive monotonicity property required of any valid measure of redundancy. We identified a set of conditions when a conditional version of Gács and Körner's common information is an ideal measure of unique information. More generally, the quest for an operationally justified decomposition of multivariate information remains an open problem. In this work, we have tried to explore the richness of this problem through the lens of network information theory. As opposed to point-to-point Shannon theory that has found extensive applications in all areas of neuroscience, we believe that the intersection between network information theory and neuroscience is virtually non-existent. Exploring the "synergy" between these two currently active research areas might provide valuable insights and possibly enrich both the fields.

**Acknowledgments**

A part of this work was supported by the Advanced VLSI Consortium, IIT Kharagpur. Thanks are due to Prof. N. B. Chakrabarti for useful discussions.